\documentclass[showpacs,twocolumn,prb]{revtex4}
\usepackage{graphicx}
\usepackage{amsmath}
\usepackage{natbib}
\usepackage{bm}
\usepackage[dvips]{color}
\usepackage{framed}
\usepackage{soul}

\begin{document}
\title{Theoretical study of $\alpha$-U/W(110) thin films from density functional theory calculations: \\
Structural, magnetic and electronic properties}
\date{\today}
\author{M. Zarshenas}
\author {S. Jalali Asadabadi}
\email[Corresponding author. Department of Physics, Faculty of Science, University
of Isfahan (UI), Hezar Gerib Avenue, Isfahan 81744, Iran. Tel.: +98 311 7934776; Fax: +98 311 7932409.\\ E-mail address: sjalali@phys.ui.ac.ir (Saeid Jalali Asadabadi).]{}
\affiliation{Department of Physics, Faculty of Science, University
of Isfahan (UI), Hezar Gerib Avenue, Isfahan 81744, Iran}

\begin{abstract}
Structural, electronic and magnetic properties were calculated for the optimized $\alpha$-U/W(110) thin films (TFs) within the density
functional theory. Our optimization for 1U/7W(110) shows that the U-W vertical interlayer spacing (ILS) is expanded by 14.0\% compared
to our calculated bulk W-W ILS. Spin and orbital magnetic moments (MMs) per U atom were found to be enhanced from zero for the bulk of $\alpha$-U to 1.4 $\mu_B$ and -0.4 $\mu_B$ at the interface of the 1U/7W(110), respectively. Inversely, our result for 3U/7W(110) TFs shows that the surface U-U ILS  is contracted by 15.7\% compared to our obtained bulk U-U spacing. The enhanced spin and orbital MMs in the 1U/7W(110) were then found to be suppressed in 3U/7W(110) to their ignorable bulk values. The calculated density of states (DOS) corroborates the enhancement and suppression of the MMs and shows that the total DOS, in agreement with experiment, is dominated in the vicinity of Fermi level by the 5f U states. Proximity and mismatch effects of the nonmagnetic W(110) substrate were assessed and found to be important for this system.
\end{abstract}



\maketitle

\section{Introduction}\label{Int}
Uranium, as the most predominant light actinide, has attracted considerable interests experimentally\cite{Iwa81, Rei85, Mol98, Mol01, Ber04, War08, Spr08} and theoretically\cite{Sod00, Jon00, Nor00, Mol01, Sto03, Ber04, Spr08} in both of its bulk\cite{Sod00, Jon00, Nor00} and surface\cite{Iwa81, Rei85, Mol98, Mol01, Sto03, Ber04, War08, Spr08} states due to the dual nature\cite{Efr04} of its 5f electrons. An experimentally detected sharp peak at 2.3 eV below the Fermi energy ($E_F$) for a thin foil of $\alpha$-U revealed that the 5f electrons could be nearly localized at the surface of this system\cite{Iwa81}. However, it was very soon realized that the peak was originated from the 6d valence band emission due to the exposure of uranium surfaces to O$_2$ which conversely implied an itinerant behavior for the 5f-electrons at the U surface\cite{Rei85}. A dispersive behavior was also
found\cite{Mol01} for the U 5f states and attributed to the direct f-f interaction, which excellently verified the band-like feature of the 5f U states reported in Ref.~\cite{Rei85}. Furthermore, the first preformed scanning tunneling spectroscopy experiment to observe 5f states\cite{Ber04} for the evaporated well ordered uranium films up to 80~\AA~on the W(110) crystal at room temperature reconfirmed that the 5f electrons would not be considered as localized electrons. Therefor, according to the aforementioned experimental evidences,\cite{Rei85, Mol01, Ber04} we do not treat $\alpha$-U/W(110) thin films (TFs) as a strongly correlated system in this work, which effectively saves the time of calculations in our ab initio surface calculations. There are several experimental evidences\cite{Mol98, Mol01, Ber04} which reveal that the structure of U overlayers is hypothetical hcp in the U/W(110) TFs. On the other hand, there are some contradicting experimental results\cite{War08, Spr08} which show that the structure of U overlayers is orthorhombic. The hcp-U TFs\cite{Mol98, Spr08} and $\alpha$-U TFs\cite{Sto03} in their hypothetical freestanding forms have been theoretically studied. The orthorhombic $\alpha$-phase is more stable than the hcp- and $\gamma$-phases\cite{Ber04}. Springell et al. within the full-potential (FP) linear muffin-tin orbital (LMTO) method calculated spin magnetic moment (MM) to be $\sim$0.1$\mu_B$\cite{Spr08} for the bulk of the $\gamma$-U compound and to be more enhanced to $\sim$0.2$\mu_B$\cite{Spr08} for the surface of the freestanding $\gamma$-U TFs. Stoji\'{c} and coworkers employing the FP-augmented plane waves (FP-APW) method have shown that the $\alpha$-U compound is a nonmagnetic system in its bulk form, whereas it behaves as a magnetic system with the imposed spin MM of 0.84$\mu_B$ at its freestanding surfaces\cite{Sto03}. A typical substrate can affect the structure\cite{Raf09} and magnetism\cite{Jal10} of the overlayers that can give rise to a variety of lattice mismatches and as a result to various U-U interlayer (IL) spacings (ILSs). The U-U spacing can change the degree of 5f localizations and thereby electronic structures. By going beyond the hypothetical freestanding hcp- and $\alpha$-U surfaces, we then aim to systematically study the surface properties and electronic structures of the realistic $\alpha$-U/W(110) TFs. On the contrary to the freestanding hcp- and $\alpha$-U TFs, our result shows that the 3U/7W(110) is not a magnetic system.

Our work is organized as follows. We first describe computational details in Sec.~\ref{Com}. Subsequently in Sec.~\ref{Opt} we elaborate on the simulation procedure. This section is presented together with the structural properties of the W and $\alpha$-U in their bulk and surface states to optimize the superlattice structure of the $\alpha$-U/W(110) TFs. In Sec.~\ref{Rel} changes of ILSs due to the relaxation in different slabs are calculated and discussed. We work out the magnetic properties by shedding light into the electronic structures of the $\alpha$-U/W(110) TFs in Sec.~\ref{MagMag}, which is divided to 5 subsections. In Sec.~\ref{Bul} we focus on the bulks of U and W as the fundamental materials of this surface study.  In Sec.~\ref{Fre} we concentrate on the hypothetical freestanding $\alpha$-U surfaces to somehow validate our subsequent predictions on the actual $\alpha$-U/W(110) thin solid films (TSFs). In this section we in excellent agreement with the reported results in Ref.~\cite{Sto03} find out that the hypothetical freestanding $\alpha$-U(001) can behave as a magnetic system. In Sec.~\ref{Sur} we shall turn our attention to $\alpha$-U/W(110) TSFs, as the main subject of this work. In Sec.~\ref{Sur} we show that the $\alpha$-U/W(110) is not a magnetic system, which is in contrast to the freestanding result. The source of such a different magnetic behavior is the subject of Sec.~\ref{Comparison} where the hypothetical freestanding $\alpha$-U(001) and the actual $\alpha$-U/W(110) TSFs are compared. In this section we concentrate on the vertical IL U-U distances as the main reason for the different behavior. We then complete the work in Sec.~\ref{Prox} with a discussion on the effects of tungsten substrate on its uranium overlayers. In this section we show the effect of tungsten on the in-plane U-U distances and as a result on the magnetization of the $\alpha$-U/W(110) system compared to that of the freestanding  $\alpha$-U(001) TFs.  Finally, we summarize and conclude the work in Sec.~\ref{Sum}.

\section{Computational Details}\label{Com}
The FP-APW + local orbitals\cite{Sjs00, Mad01} (FP-APW+lo) method is used for solving the set of density functional theory (DFT)  equations\cite{Hoh64,Koh65} as embodied in the WIEN2k code\cite{Bla01} to solve the Kohn-Sham equations\cite{Koh65}. For the exchange-correlation potential no more correlations are included than what have been implemented in the Perdew and Wang local density approximation\cite{Per92} (PW-LDA) and the Perdew-Burke-Ernzerhof generalized gradient approximation\cite{Per96} (PBE-GGA), because of the reported band-like character in Refs.~\cite{Rei85, Mol01, Mol98, Ber04, War08, Spr08} for the 5f U electrons in U/W(110) system. The orbital polarization has been estimated using a non-self-consistent treatment, which may be reasonable due to the assumed itinerant character for the 5f states. The relativistic spin-orbit (SO) interactions were incorporated to the band structure calculations in a second variational procedure. The Muffin-tin radii were adjusted to R$_{MT}$=2.4 bohr for the W and to R$_{MT}$=2.5 bohr for the U atoms. The expansion of the wave functions and charge densities were cut off by the R$_{MT}$K$_{max}$ = 7 and G$_{max}$ = 12 parameters, respectively. The full relaxations were performed with the criterion of 1 mRy/bohr on the exerted forces. A set of
14$\times$14$\times$1 special k-points has been used for integrations over the Brillouin zone of the 1$\times$1 surface cell.

\section{Optimization of Superlattice Surface Structure} \label{Opt}

\begin{figure}
\includegraphics[width=9.0cm]{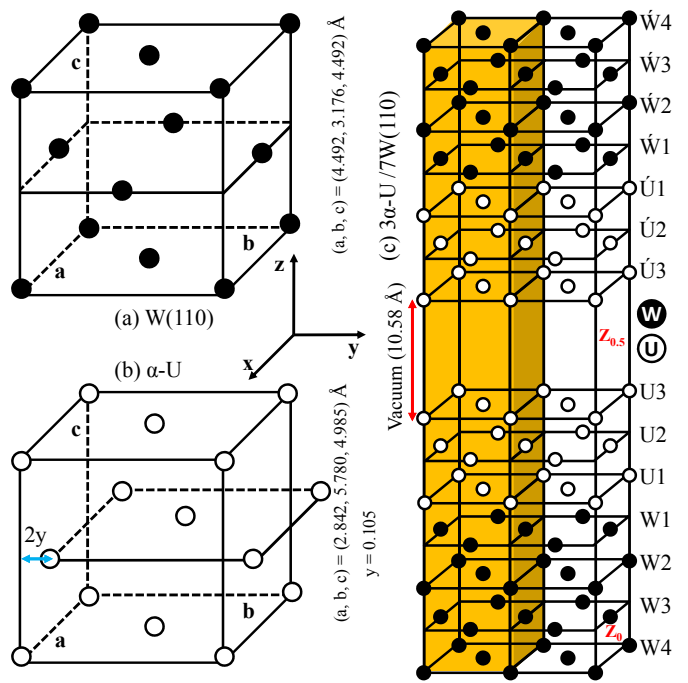}
\caption{(Color online) Black and white circles represent tungsten
(W) and uranium (U) atoms, respectively. (a) Crossview of
bcc-tungsten along [110] direction as our clean W(110) substrate
with a = c = $\sqrt{2}~$a$_{bulk}$ = 4.492~\AA~and b = b$_{bulk}$
= 3.176~\AA. (b) Orthorhombic structure of $\alpha$-U with a =
2.842~\AA, b = 5.780~\AA, c = 4.985~\AA~and the relaxed internal
parameter y = 0.105. (c) Doubled superlattice structure of
3U/7W(110) TFs by depositing $\alpha$-U after shifting its
middle plane to the right-hand side by the (1/2 - 2y)\textbf{b} on
W(110) taking lattice mismatch and enough amount of symmetric
vacuum, 20 bohr = 10.58~\AA, into account. Left highlighted
portion constitutes the conventional unit cell of our
$\alpha$-U/7W slab. The primitive unit cell of the highlighted
slab constitutes our superlattice structure. The super lattice
structure crystallizes in the Cmmm space group and contains
totally 13 atoms, 2$\times$3=6 U atoms and 2$\times$3+1=7 W atoms.
\'{U}(\'{W}) represents image of U(W) with respect to an imaginary
central plane which horizontally crosses the slab. Z$_0$
(Z$_{0.5}$) shows the midpoint of the substrate (vacuum) of the
3U/7W(110) system.} \label{fig1}
\end{figure}

We calculated the lattice parameter of the bcc-W to be 3.154~\AA~within LDA and 3.176~\AA~within GGA. These results are in agreement with the pseudopotential results of the others\cite{Kuc10}, i.e., 3.129~\AA~ within LDA and 3.175~\AA~within GGA. The GGA results are in more agreement with the experimental value\cite{Gaf91} of  3.165~\AA~than those of the LDA. We cut the bcc unit cell of the tungsten by the (110) plane and reconstructed another unit cell along the [110] crystallography direction. As shown in Fig.~\ref{fig1}~(a), the reconstructed unit cell as our clean W(110) substrate looks like the fcc structure with a = c $\neq$ b. Lattice parameters of the fcc-like W(110) structure were fixed to be a = c = $\sqrt{2}~$a$_{bulk}$ = 4.492~\AA~and b = b$_{bulk}$ = 3.176~\AA, where our GGA lattice parameter for the bcc-W, 3.176~\AA, was used for a$_{bulk}$ and b$_{bulk}$. Seven monolayer (ML) W(110) films were found quite enough to model the substrate. We fully relaxed the orthorhombic $\alpha$-U, as shown in Fig.~\ref{fig1}~(b), and calculated the lattice parameters including internal parameter y to be (a, b, c, y) =
(2.842~\AA, 5.780~\AA, 4.985~\AA, 0.105) in agreement with the experimental values\cite{Bar63} of (a, b, c, y) = (2.844~\AA, 5.869~\AA, 4.932~\AA, 0.102). Our calculated equilibrium volume of the primitive unit cell for the $\alpha$-U, V$_{U_\alpha}$ = abc/4 = 20.472~\AA$^3$, is in excellent agreement with the experimental value\cite{Bar63}, 82.3256/4 = 20.581~\AA$^3$, and can be compared with the values obtained from other theoretical methods\cite{Sod00, Jon00, Nor00} as well. The $\alpha$-U structure represented in Fig.~\ref{fig1}~(b) can be transformed to the fcc-like structure, if its middle horizontal plane is shifted to the right-hand side along y-axis by the (1/2 - 2y)\textbf{b}. In this case, the Fig.~\ref{fig1}~(b) looks like Fig.~\ref{fig1}~(a), as both of them are now fcc-like. However, their lattice parameters are not identical to each other, as discussed above and shown in Figs.~\ref{fig1}. Therefore, the substrate, due to its different in-plane and vertical lattice parameters compared to those of the U overlayers, gives rise to lattice mismatches. By considering the in-plane lattice parameters, a = 4.492~\AA, b = 3.176~\AA, and vertical ILS,  lattice parameter, c/2 = 4.492/2~\AA~= 2.246~\AA, of the clean W(110) surface, 2$\times$3 ML of the $\alpha$-U were symmetrically deposited on top and bottom of the 7 ML W(110). The three U layers on bottom of the 7W(110), i.e. \'{U1}, \'{U2}, \'{U3}, which can be considered as the images of those on top of the substrate, were also deposited to add inversion symmetry. In order to reach the highest possible symmetry, less traditional slab is set up. The vacuum is then added in the interior of the slab, as shown in Fig.~\ref{fig1}~(c). In traditional slab the vacuum is usually added on top and below of the slab. Here, in contrast to the traditional slab, the vacuum is created in between of the outer U layer (U3) and its image \'{U3}, as shown in Fig.~\ref{fig1}~(c). The \'{Wi} also represents the image of Wi with respect to an imaginary central plane which horizontally crosses the slab for i=1, 2, 3, 4. The vacuum thickness was determined to be 10.58~\AA~by calculating the total energy, work function, and exerted forces on the surface atoms versus various vacuum thicknesses for the 7 ML W(110) clean substrate. The total energy, work function and forces were well converged at the determined vacuum thickness - they did not considerably fluctuate by increasing the vacuum thickness more than the value obtained. The jU/iW(110) layers were individually fully relaxed for i = 1 to 4 and j = 0 to 3; j = 0 refers to the uncovered iW(110). The conventional unit cell of the 3U/7W(110) system is highlighted, as shown in Fig.~\ref{fig1}~(c). The primitive unit cell of the highlighted portion of the relaxed slab, which crystallizes in the \emph{Cmmm} space group,  constitutes our superlattice surface structure.

\section{Relaxed Interlayer Spacings}\label{Rel}
\begin{figure}
\includegraphics[width=9.0cm]{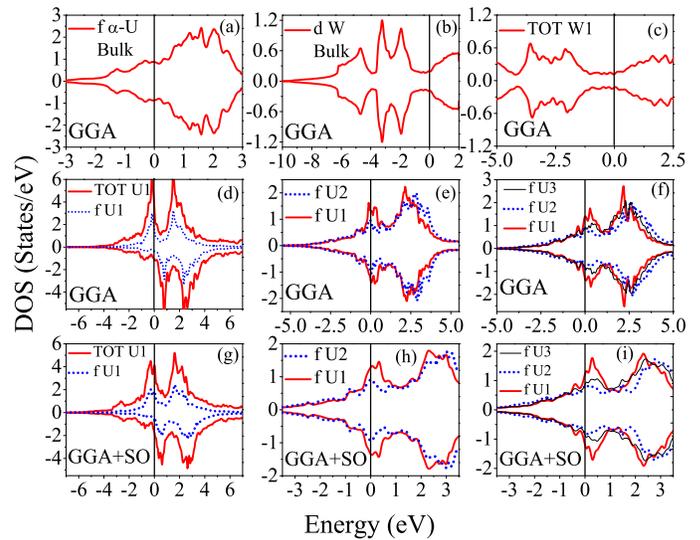}
\caption{(Color online) DOSs  for (a) $\alpha$-U compound, (b)
bcc-W compound, (c) 1U/7W(110), (d) 1U/7W(110), (e) 2U/7W(110),
(f) 3U/7W(110), (g) 1U/7W(110), (h) 2U/7W(110), (i) 3U/7W(110).}
\label{fig2}
\end{figure}

\begin{table*}[!t]
 \begin{center}
 \caption{Vertical ILDs for fully relaxed covered and uncovered tungsten substrate surfaces and subsurfaces in \AA. The percentages of the relative differences given in columns 2 and 3 of  the table were obtained by comparing with the bulk U-U vertical ILD of d$_{U-U}^{bulk}$(001) = 2.493~\AA, while for the remained columns 4, 5, 6, and 7, they were obtained by comparing with the bulk W-W vertical ILD of d$_{W-W}^{bulk}$(110) = 2.246~\AA.}\label{tab1}
  \begin{ruledtabular}
   \begin{tabular}{lccccccc}
      System          &d$^{sur}_{U3-U2}$&d$^{sur}_{U2-U1}$&d$^{sur}_{U1-W1}$&d$^{sur}_{W1-W2}$&d$^{sur}_{W2-W3}$&d$^{sur}_{W3-W4}$&\\
      \hline
      1U/7W(110)      &&&2.562 (14.0\%)&2.202 (-1.9\%)&2.196 (-2.2\%)&2.208 (-1.7\%)& \\
      2U/7W(110)      &&2.018 (-19.1\%) &2.397 (6.7\%)&2.116 (-5.8\%)&2.114 (-5.9\%)&2.114 (-5.9\%)&\\
      3U/7W(110)      &2.102 (-15.7\%)&2.164 (-13.2\%)&2.432 (8.3\%)&2.149 (-4.3\%)&2.143 (-4.6\%)&2.140 (-4.7\%)&\\
      5W(110)         &&&&2.124 (-5.3\%)&2.226 (-0.9\%)&&\\
      $^{a}$5W(110)   &&&&~~~~~~~(-4.1\%)&~~~~~~~(-0.2\%)&&\\
      $^{b}$5W(110)   &&&&2.173 (-4.1\%)&2.258 (-0.4\%)&&\\
      7W(110)         &&&&2.140 (-4.7\%)&2.208 (-1.7\%)&2.219 (-1.2\%)&\\
      $^{a}$7W(110)   &&&&~~~~~~~(-3.3\%)&~~~~~~~(-0.1\%)&~~~~~~~(-0.4\%)&\\
      $^{c}$12W(110)  &&&&~~~~~~~(-4.7\%)&~~~~~~~(0.2\%)&~~~~~~~(-0.6\%)&\\
      $^{a}$W(110)$^{Exp.}$    &&&&(-3.1\%$\pm$0.6\%)&(0.0\%$\pm$0.9\%)&(0.0\%$\pm$1.0\%)&\\
   \end{tabular}
  \end{ruledtabular}
 \end{center}
 \begin{flushleft}
$^a$Reference~\textrm{\cite{Arn97}; The theoretical data were
calculated within the APW method as embodied in the WIEN93 code
using the PZ-LDA81. The experimental data were obtained from LEED
analysis.}\\
$^b$Reference~\textrm{\cite{Qia99}}; The data were calculated
within the FP-APW method as embodied in the WIEN97 code using the
GGA92.\\
$^c$Reference~\textrm{\cite{Spi04}}; The data were calculated
within the pseudopotential method as embodied in the Vasp code using
the GGA92.\\
\end{flushleft}
\end{table*}

Vertical IL lattice spacings were calculated for the covered and uncovered W(110) surfaces and subsurfaces. The results after relaxations together with the experimental\cite{Arn97} and theoretical\cite{Arn97, Qia99, Spi04} results are presented in Tab.~\ref{tab1}. The relative distances in percentage were also given in this table. The relative changes for the two last shallow uranium layers were obtained by comparing with the bulk U-U vertical IL distance (ILD) of d$_{U-U}^{bulk}$(001) = 2.493~\AA. The results are presented in columns 2 and 3 of Tab.~\ref{tab1}. For the first deposited uranium layer and all of the tungsten layers, as presented in columns 4 to 7 of Tab.~\ref{tab1}, the relative changes were obtained by comparing with the bulk W-W vertical ILD of d$_{W-W}^{bulk}$(110) = 2.246~\AA. For 7 ML W(110) clean substrate, our result shows that the surface shallow vertical ILS d$_{W1-W2}^{sur}$(110) is contracted to 2.140~\AA by 4.7\%,  which is in agreement with the experimental value\cite{Arn97} of (-3.1$\pm$0.6)\% and theoretical value\cite{Arn97} of -3.3\%. Our calculated contraction d$_{W1-W2}^{sur}$(110) result for 5 ML W(110) clean system, 5.3\%, is in agreement with the result of the others\cite{Arn97}, 4.1\%, for the same system.  Our 5W(110) and 7W(110) results show that the shallow ILS by adding one more bilayers is expanded from -5.3\% for 5W(110) to -4.7\% for 7W(110). The same trend was reported earlier,\cite{Arn97} as shown in Tab.~\ref{tab1}. However, this trend cannot be generalized to every layers, as a small expansion, which is not presented here, is occurred by going from 7W(110) to 9W(110) in agreement with Ref.~\cite{Arn97}. The accuracy of the experimental results, as represented in Tab.~\ref{tab1}, is about 1.0\% for deeper layers of tungsten which are given in the last two columns of Tab.~\ref{tab1}. For the later layers our results in agreement slightly  with the theoretical results of the others\cite{Arn97, Qia99, Spi04}, predict small relaxations near the range of the experimental accuracy. The small relative changes can be then nearly neglected for these deeper layers of W(110). Now, let us gradually add the three uranium adatoms on the 7W(110) substrate layer by layer. By adding the first ML of uranium on the 7W(110) substrate (ultrathin 1U/7W(110) film), the result, as presented in Tab.~\ref{tab1}, shows that the vertical ILS at the interface, d$^{sur}_{U1-W1}$, is considerably expanded by 14\% to 2.562~\AA~when compared with the vertical ILS of the bcc-bulk along the [110] direction, i.e., d$_{W-W}^{bulk}$(110) = 2.246~\AA. Because of the large expansion, one may expect to observe significant changes at the interface in the electronic structures and as a result in the sensitive  physical properties, as shall be discussed in Sec.~\ref{MagMag} and subsections therein . Conversely, for the 2U/7W(110) system a dramatic contraction, 19.1\%, is observed in Tab.~\ref{tab1} for the outer vertical ILS of d$^{sur}_{U2-U1}$ = 2.018~\AA~compared to the bulk of uranium ILS along the [001] direction, i.e., d$_{U-U}^{bulk}$(001) = 2.493~\AA. The contraction gives rise to an enhancement of the direct 5f-5f interactions between two adjacent uranium atoms along the z-Cartesian direction. Consequently, one may expect to observe more broadened density of states (DOS) for the 2U/7W(110) system compared to the 1U/W(110) system, as shall be shown in Figs.~\ref{fig2}. By adding the third uranium layer to the last 2U/7W(110) slab and formation of the 3U/7W(110) system, the new outer ILD, d$^{sur}_{U3-U2}$ = 2.102~\AA, is slightly expanded compared to the last outer ILS of d$^{sur}_{U2-U1}$ = 2.018~\AA. But, the later distance of d$^{sur}_{U3-U2}$ is still contracted by 15.7\% compared to the d$_{U-U}^{bulk}$(001) = 2.493~\AA~and by 19.8\% compared to the outer ILS of d$^{sur}_{U1-W1}$ = 2.562~\AA~for the ultrathin 1U/7W(110) film. Therefore, one may expect to observe similar behavior for both of 3U/7W(110) and 2U/7W(110) systems. But these 3U/7W(110) and 2U/7W(110) systems may be expected to behave differently compared to the ultrathin 1U/7W(110) film, as shall be shown in  Secs.~\ref{Fre},~\ref{Comparison}~and~\ref{Prox}. As can be seen from columns 4 to 7 of Tab.~\ref{tab1}, our result shows that the ILSs oscillate versus uranium film thickness. For instance, the d$^{sur}_{W1-W2}$ first decreases from 1U/7W(110) to 2U/7W(110), while it then increases from 2U/7W(110) to 3U/7W(110). The later oscillation can be also seen from the  d$^{sur}_{U1-W1}$, d$^{sur}_{W2-W3}$, and d$^{sur}_{W3-W4}$ columns as well as the aforesaid d$^{sur}_{W1-W2}$.

\section{Magnetic Properties and Electronic Structures}\label{MagMag}
\subsection{Bulks of U and W}\label{Bul}
We calculated the MMs and the DOS for the orthorhombic $\alpha$-U. The DOS showed in Fig.~\ref{fig2}~(a) indicates that the system is nonmagnetic in its bulk form. This result is in excellent agreement with the previous results of the others\cite{Sto03, Spr08} within different methods. It was shown that the bulk of $\alpha$-U compound could be considered as a nonmagnetic system using FP-APW method in Ref.~\cite{Sto03} and using FP-LMTO method in Ref.~\cite{Spr08}. The later group\cite{Spr08} calculated spin and orbital MMs per uranium atom to be 0.1 $\mu_B$ and -0.22 $\mu_B$  for the hypothetical hcp-U in its bulk form, respectively. This shows that the structure of the uranium can affect the magnetic properties of the uranium based systems. Thus, it may be important to take an actual structure into consideration for the uranium as the overlayer deposited on the tungsten substrate. We also made sure about the magnetism of the tungsten substrate in its bulk form. The bulk of tungsten is also found to be nonmagnetic, as may be seen from Fig.~\ref{fig2}~(b).
\subsection{Freestanding U Surfaces}\label{Fre}
\begin{figure}
\includegraphics[width=5.0cm]{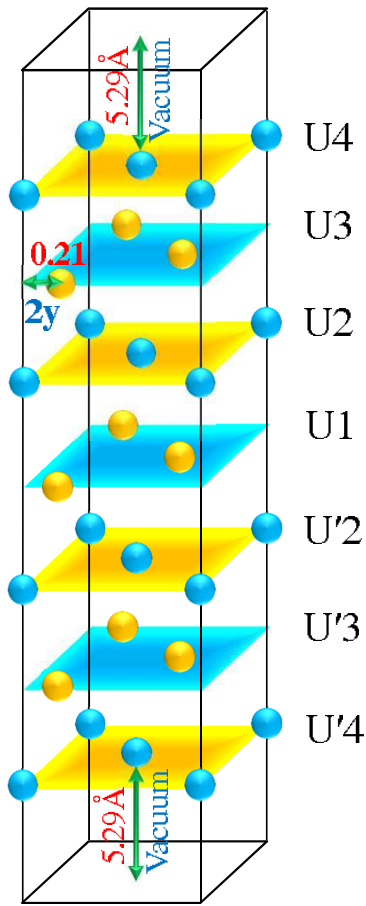}
\caption{(Color online) Freestanding 7 ML $\alpha$-U(001).}
\label{fig3}
\end{figure}

\begin{table}
 \begin{center}
  \caption{Spin MM inside the Muffin-tin sphere (MT), spin MM in the interstitial region (Int), total spin MM per U atom(SP), orbital MM (Orb) and total MM (Tot) in $\mu_B$ within GGA+SO for the freestanding $\alpha$-U(001) TFs from 1 ML to 7 ML. For more than 1 layer the total spin MMs per atom were calculated as MT+$\frac{1}{2}$Int.}\label{tab2}
  \begin{ruledtabular}
   \begin{tabular}{lccccccc}
                      &1 layer&3 layers&5 layers&7 layers\\
     \hline
      MT         &2.132&0.867&0.864&0.865\\
      Int.      &0.754&0.412&0.409&0.411\\
      SP       &2.886&1.073&1.068&1.070\\
      Orb     &-0.812&-0.372&-0.372&-0.373\\
      Tot
                &2.074&0.701&0.696&0.697\\
               \end{tabular}
  \end{ruledtabular}
 \end{center}
\end{table}

To reinforce our approach and validate its predictions on the actual $\alpha$-U/W(110) case, it may be expedient to first try to reproduce the freestanding results. The freestanding results can be directly compared with the theoretical results of the others. We constructed a slab containing 7 ML U, as illustrated in Fig.~\ref{fig3}, to simulate the freestanding $\alpha$-U. We increased the vacuum to reach the optimized value of 20 bohr, i.e. $\sim$5.29$\times$2 = 10.58\AA, as shown in Fig.~\ref{fig3}. We made sure that total energy, surface energy and work function of the system were not changed by increasing the amount of vacuum. We then fully relaxed the uranium layers to minimize the exerted forces on each atoms, as mentioned in Sec.~\ref{Com}. We calculated spin MM inside the Muffin-tin sphere, total spin MM, orbital MM inside the Muffin-tin sphere, interstitial and total MMs within the GGA+SO for the freestanding $\alpha$-U surfaces. Our results are presented in Tab.~\ref{tab2}. The total spin MM can be obtained by adding the spin MM inside the Muffin-tin sphere to its interstitial contribution. Stoji\'{c} and collaborators\cite{Sto03} calculated spin MM inside the Muffin-tin sphere, orbital MM inside the Muffin-tin sphere, interstitial, and total MMs per U atom to be 2.11 (0.84), -0.83 (-0.38), 0.74 (0.39), and 2.02 (0.66) in $\mu_B$ for 1 ML (3 ML) of the hypothetical freestanding $\alpha$-U TFs using the FP-APW method within the GGA+SO, respectively. Their results\cite{Sto03} were well converged and did not show significant changes by adding more uranium overlayers. A comparison shows that our freestanding results, as presented in Tab.~\ref{tab2}, are in good agreement with the results given in Ref.~\cite{Sto03}. The small differences between our results and the results reported in Ref.~\cite{Sto03} could be due to the fact that they\cite{Sto03} did not relax their simulated freestanding $\alpha$-U TFs and utilized the unrelaxed experimental geometry. The agreement authenticates the validity of our freestanding results. Spin and orbital MMs per uranium atom were calculated in Ref.~\cite{Spr08} to be $\sim$0.2 $\mu_B$ and $\sim$-0.44 $\mu_B$ for the hypothetical freestanding hcp-U surface, respectively. Our result shows that the spin and orbital MMs of the first uranium overlayer are aligned antiparallel in accordance with the above results reported in Refs.~\cite{Sto03, Spr08}. Spin MM and the magnitude of the orbital MM of $\alpha$-U freestanding TFs are larger than those of the hcp-U freestanding TFs. This is in the case that, as discussed in Sec.~\ref{Bul}, the hcp-U shows magnetic ordering in its bulk form, whereas the bulk of $\alpha$-U is not a magnetic system. Hence, the structure of uranium can cause to reverse the magnetic properties from its bulk to its surfaces and inversely from its surfaces to its bulk. Similar results were recently reported\cite{Jal10} for the isostructural $\alpha$- and $\gamma$-Ce systems.
\subsection{$\alpha$-U/W(110) Thin Films}\label{Sur}
\begin{table}
 \begin{center}
 \caption{Spin MM inside the Muffin-tin sphere of Uj atom (MT(Uj) for j=1 to $j_{max}$), and of Wi atom (MT(Wi) for i=1 to 4), spin MM in the interstitial region (Int.), total spin  MM of the primitive unit cell (SP), orbital MM inside
 the Muffin-tin sphere of Uj atom (ORB(Uj) for j=1 to $j_{max}$), and total MM per primitive unit cell (TOT) in $\mu_B$
 in covered jU/7W(110) for j = 1 to $j_{max}$ within the PW-LDA and PBE-GGA with and without SO coupling. The total spin moment is SP=2$\times(\sum_{j=1}^{j_{max}}$MT(Uj)+$\sum_{i=1}^4$MT(Wi))+Int. Total MM per primitive unit cell (TOT) is TOT=SP+2$\times\sum_{j=1}^{j_{max}}$ORB(Uj). The $j_{max}$ index varies from 1 in 1U/7W to 3 in 3/7W(110).}\label{tab3}
   \begin{ruledtabular}
   \begin{tabular}{lcccccc}
    &GGA&GGA+SO&LDA&LDA+SO  \\
    \hline
    &&1U/7W(110)\\
    MT(U1)       &1.707&1.402&1.445&1.259  \\
    MT(W1)       &0.000&-0.011&0.009&0.017  \\
    MT(W2)       &0.003&0.010&-0.003&0.001 \\
    MT(W3)       &-0.003&-0.002&-0.006&-0.001  \\
    MT(W4)       &-0.003&-0.003&-0.001&0.003  \\
    Int.         &1.150&1.088&0.871&0.825  \\
    SP           &4.558&3.880&3.759&3.383\\
    ORB(U1)      &&-0.446&&-0.386  \\\hline
    TOT          &4.558&2.988&3.759&2.611  \\\hline
    &&2U/7W(110)\\
    MT(U2)       &0.126&-0.030&0.000&0.000  \\
    MT(U1)       &0.040&-0.008&0.000&0.000\\
    MT(W1)       &0.002&0.000&0.000&0.000  \\
    MT(W2)       &0.002&0.000&0.000&0.000 \\
    MT(W3)       &0.000&0.000&0.000&0.000  \\
    MT(W4)       &-0.001&0.000&0.000&0.000  \\
    Int.         &0.109&-0.031&0.001&0.000  \\
    SP           &0.447&-0.107&0.001&0.000\\
    ORB(U2)      &&0.008&&0.000 \\
    ORB(U1)      &&0.002&&0.000 \\\hline
    TOT          &0.447&0.057&0.001&0.000  \\\hline
    &&3U/7W(110)\\
    MT(U3)       &0.086&-0.021&-0.022&0.009  \\
    MT(U2)       &0.020&-0.003&-0.007&0.000\\
    MT(U1)       &0.010&0.000&0.002&0.001\\
    MT(W1)       &-0.002&0.000&0.000&0.000  \\
    MT(W2)       &0.000&0.000&0.000&0.000 \\
    MT(W3)       &-0.001&0.000&0.000&0.000  \\
    MT(W4)       &0.000&0.000&0.000&0.000  \\\
    Int.         &0.050&-0.013&-0.011&0.001  \\
    SP           &0.276&-0.061&-0.065&0.020\\
    ORB(U3)      &&0.006&&-0.002 \\
    ORB(U2)      &&0.001&&0.000 \\
    ORB(U1)      &&0.000&&0.000 \\\hline
    TOT          &0.276&-0.047&-0.055&0.016  \\
   \end{tabular}
  \end{ruledtabular}
 \end{center}
\end{table}

Spin MMs per atom were calculated for the U and W in the jU/7W(110) system with j = 1, 2, 3 using the LDA and GGA excluding and including SO coupling. Orbital polarizations were included only for the U atoms in a non-self-consistent manner. The non-self-consistent procedure seems to be adequate for predicting orbital moment, because 5f U electrons in this system behave as band-like electrons. To obtain self-consistently the orbital MMs of the U overlayers, one may include suitable correlations among 5f-electrons, e.g., by performing LDA+U calculations. Here, instead, following the experimental results\cite{Rei85, Mol01, Mol98, Ber04} we assumed that the 5f electrons were not localized in U/W(110) TFs, as notified in Sec.~\ref{Com}. The results together with the total MM per primitive unit cell are presented in Tab.~\ref{tab3}. One may observe some palpable trends which can be immediately deduced from Tab.~\ref{tab3}. The magnitude of our calculated MMs within the GGA is larger than that of the LDA. We can come to the same conclusion by comparing the result of the GGA+SO with that of the LDA+SO. This implies that the SO coupling does not affect the trend. However, SO gives rise to reduce the MMs as can be observed by comparing the result of LDA (GGA) with that of the LDA+SO (GGA+SO), as presented in Tab.~\ref{tab3}. The inclusion of the SO coupling makes the up-DOS more symmetric with respect to the down-DOS compared to the SO excluded case (see  Figs.~\ref{fig2}~(d), and (g)). Thereby, the difference between up- and down-DOSs becomes smaller by including SO coupling due to the more symmetric up- and down-DOSs. The more symmetric up- and down-DOSs may result in less MM. Therefore, one may expect to obtain smaller MM by including SO coupling for this system. The reduction of MM from GGA to GGA+SO is comparable with the reduction of MM from GGA to LDA, as shown in Tab.~\ref{tab3}. Therefore, the results of GGA+SO and LDA seem to be approximately closer to each other than the results of the other exchange-correlation functionals (see Tab.~\ref{tab3}). Our surface MM calculation does not reveal any considerable MMs for Wi with i = 1 to 4. Total up-DOS of W1 cancels total down-DOS of W1 (see Fig.~\ref{fig2}~(c)). Therefore, our result, in agreement with Ref.~\cite{Hua05}, shows that the tungsten even at its surfaces is not capable of producing magnetism. Our result predicts spin MM at the first layer of uranium (U1) for the ultrathin 1U/7W(110) system, as shown in Tab.~\ref{tab3}. Orbital MM is also predicted by our calculations including SO coupling in 1U/7W(110). The total- and f-DOSs for U1 in ultrathin 1U/7W(110) film, as shown in Fig.~\ref{fig2}~(d), corroborate that the nonmagnetic bulk $\alpha$-U becomes a magnetic layer when its first ultrathin layer is deposited on the nonmagnetic W(110) substrate. The source of the imposed MM in the first layer of 1U/7W(110) may be attributed to the vertical ILD at the interface, as presented in Tab.~\ref{tab1}. The result shows that the U1-W1 distance, d$^{sur}_{U1-W1}$ = 2.256~\AA, is expanded by 14\% compared to d$^{bulk}_{W-W}(100)$ = 2.246~\AA, and by 2.8\% compared to d$^{bulk}_{U-U}(100)$ = 2.493~\AA. More importantly, there is no any other uranium layer over the first deposited U layer in the case of 1U/7W(110). Thus, one does not account for the strong direct 5f-5f interactions perpendicular to the surface. This gives rise to a more localized DOS near the $E_F$, as shown in Fig.~\ref{fig2}~(d). Thereby, there is no surprise in observing magnetism in the ultrathin 1U/7W or in the freestanding 1ML U system.
Our result, as shown in Tab.~\ref{tab3}, shows that by adding the second U layer to the 1U/7W(110) or by adding the third layer to the 2U/7W(110), all the components of the total MMs are substantially reduced. These sensible reductions in the MM may be attributed to the U-U ILSs in the 2U/7W(110) or 3U/7W(110), which are smaller than the Hill limit\cite{Hil70}. According to the Hill criteria, uranium based compounds with the U-U interatomic spacings smaller than 3.4-3.6~\AA~are typically nonmagnetic with itinerant 5f states\cite{Hil70}. Therefore, 2U/7W(110) with d$^{sur}_{U2-U1}$= 2.018~\AA~ and 3U/7W(110) with d$^{sur}_{U2-U1}$= 2.164~\AA~ and d$^{sur}_{U3-U2}$= 2.102~\AA, would be almost nonmagnetic based on the Hill limit. Our LDA+SO and nearly LDA as well as GGA+SO results show that the 2U/W(110) and 3U/W(110) systems are not magnetic, as shown in Tab.~\ref{tab2}, in agreement with the Hill prediction. Our calculated DOSs, as shown in the Figs.~\ref{fig2}~(e), (f), (h) and (i), authenticate that the 2U/7W(110) and/or 3U/7W(110) TFs become nonmagnetic compared to the magnetic 1U/7W(110) ultrathin film, as shown in the Figs.~\ref{fig2}~(d) and (g), by absorbing additional uranium overlayers.

\subsection{Comparison of freestanding $\alpha$-U(001) and $\alpha$-U/W(110) Thin Films}\label{Comparison}
\begin{table}[!t]
 \begin{center}
  \caption{Vertical ILDs for fully relaxed 3 ML, 5 ML and 7 ML freestanding $\alpha$-U in
 \AA.} \label{tab4}
  \begin{ruledtabular}
   \begin{tabular}{lccccccc}
      System         &d$^{sur}_{U2-U1}$&d$^{sur}_{U3-U2}$&d$^{sur}_{U4-U3}$&d$^{sur}_{U'2-U1}$&d$^{sur}_{U'3-U'2}$&d$^{sur}_{U'4-U'3}$&\\
      \hline
      3U(001)      &2.404&&&2.405&&& \\
      5U(001)      &2.340&2.275&&2.342&2.278&&\\
      7U(001)      &2.325&2.261&2.187&2.326&2.263&2.188&\\
   \end{tabular}
  \end{ruledtabular}
 \end{center}
\end{table}

The calculated spin MM inside the Muffin-tin sphere of the 1U/7W(110), 1.402 $\mu_B$, as given in Tab.~\ref{tab3}, is smaller than the value of 2.1 $\mu_B$ for the freestanding 1 ML $\alpha$-U, as presented in Tab.~\ref{tab2} or in Ref.~\cite{Sto03}. If we compare the interstitial MM of the 1U/7W(110), 1.088 $\mu_B$, as given in Tab.~\ref{tab3}, with that of 0.75 or 0.74 $\mu_B$ for the freestanding 1 ML $\alpha$-U, as presented in Tab.~\ref{tab2} or Ref.~\cite{Sto03}, we should first divide the one for 1U/7W(110) by 2, i.e., 1.088/2 = 0.544 $\mu_B$. In our 1U/7W case for 1 ML U there are two U atoms, which is not the case for the 1 ML of the freestanding $\alpha$-U system. For the later case, 1 ML freestanding $\alpha$-U, there is only one U atom in the supercell. The comparison shows that the value of 0.74 $\mu_B$ for the freestanding $\alpha$-U is larger than the calculated value of 0.544 $\mu_B$ for the 1 ML $\alpha$-U deposited on the W(110) substrate. The difference may be due to the proximity effect of the substrate. All of the atoms in the superlattice structure contribute to the interstitial MM, and there at the interstitial region their contributions are mixed with each other. However, similar to comparing total MM it is enough to divide the interstitial moments by 2, as the W atoms do not contribute substantially in the total MM. The comparison shows that the  GGA+SO total MM per U atom, $\sim$1.402+1.088/2-0.446=1.5 $\mu_B$, as given in Tab.~\ref{tab3}, regardless of the ignorable W atoms' contributions, is less than the freestanding 1 ML total MM of 2.074 or 2.02 $\mu_B$, as presented in Tab.~\ref{tab2} or in Ref.~\cite{Sto03}. This may demonstrate the effect of substrate. To reconfirm this fact, we can compare the orbital MMs as well. Even though the signs of the orbital MM for the 1U/7W(110) and the freestanding $\alpha$-U are the same, the magnitude of the GGA+SO result for the 1U/7W(110), 0.446 $\mu_B$, is half of the value for 1 ML freestanding $\alpha$-U, 0.8 $\mu_B$. This shows that all the components of the total MM for the 1U/7W(110) are reduced when compared with those of the freestanding 1 ML $\alpha$-U. The reduction may come from the nearness effect of substrate on the TF magnetism. In addition to the discussed comparison between 1 ML freestanding U and the 1U/7W(110), one may compare the trend of the MMs for the freestanding $\alpha$-U by adding more uranium layers with that of the $\alpha$U/7W(110).
A significant reduction in total MM from $\sim$2.0 $\mu_B$ for 1 ML to $\sim$0.7 $\mu_B$ for 3 ML can be observed in the freestanding $\alpha$-U TFs, as presented in Tab.~\ref{tab2} or in Ref.~\cite{Sto03}. This shows that the number of layers can be of significant importance. However, we still observe the nonzero total MM of $\sim$0.7 $\mu_B$ for 3 and larger ML of the freestanding $\alpha$-U, as presented in Tab.~\ref{tab2} or in Ref.~\cite{Sto03}. Based on the hypothetical freestanding $\alpha$-U model in agreement with Ref.~\cite{Sto03}, one predicts the surface magnetism. Our GGA+SO result predicts small MMs of 0.057 $\mu_B$ and -0.047 $\mu_B$ per primitive unit cell for the 2U/7W(110) containing 4 U atoms and 3U/7W(110) containing 6 U atoms, respectively. Therefore, the maximum total MM per U atom that can be predicted by our GGA+SO result is only 0.0014 $\mu_B$ for 2U/7W(110) and -0.0078  $\mu_B$ for 3U/7W(110) TFs. The later tiny value of 0.0014 $\mu_B$ per U for 3U/7W(110) is much smaller than the value of $\sim$0.7$\mu_B$ per U for the freestanding 3 ML $\alpha$-U. In essence, our result based on the actual $\alpha$-U/W(110) system predicts that the 3U/W(110) is not a magnetic system, while the result for the freestanding case shows that the 3 ML $\alpha$-U is a magnetic system.                                                          In order to find the physics behind the later result, we calculated the vertical IL lattice spacings for the freestanding surfaces, as shown in Fig.~\ref{fig3}. The results after relaxations were presented in Tab.~\ref{tab4}. The vertical ILD before relaxation, as shown in Fig.~\ref{fig1} (b), is the distance between the uranium layers of the $\alpha$-U in its bulk form along the [001] direction, i.e., d$_{U-U}^{bulk}$(001) = 2.493~\AA. Although the result, as can be seen from Tab.~\ref{tab4}, shows that the vertical ILSs for the relaxed freestanding surfaces, as shown in Fig.~\ref{fig3}, are contracted by relaxation compared to the unrelaxed experimental geometry, as shown Fig.~\ref{fig1} (b), the vertical ILSs are still larger than those for the $\alpha$-U/W(110) TFs, as can be seen from Tab.~\ref{tab1}. For instance, the vertical ILSs are 2.404 and 2.405~\AA~for the freestanding 3 ML $\alpha$-U(001) TFs, which are larger than those of 2.102 and 2.164~\AA~for the 3U/7W(110), as can be seen from Tabs.~\ref{tab1} and ~\ref{tab4}. The smaller U-U spacings, the stronger U-U interactions among their 5f electrons. The stronger 5f U-5f U interaction gives rise to more broadened 5f DOS, which results in less localization. Therefore, one may expect to obtain less MM for the 3U/7W(110) system compared to the freestanding 3 ML $\alpha$-U(001) TFs. Our result verifies  such an expectation, as shown in Tabs.~\ref{tab1} and ~\ref{tab4}. Our result, as shown in Tab.~\ref{tab4}, shows that by adding more layers, the vertical IL U-U spacings are reduced for the freestanding U TFs. For example, the distance between U1 and U2 is reduced from 2.404~\AA~for 3 ML freestanding U(001) to 2.340~\AA~for 5 ML freestanding U(001), as shown in Tab.~\ref{tab4}. This reduction may explain why the MM is reduced from 0.701 $\mu_B$ for 3 ML freestanding U(001) to 0.696 $\mu_B$ for 5 ML freestanding U(001).
\subsection{Proximity and lattice mismatch effects of W(110)
substrate}\label{Prox}
\begin{figure*}
\includegraphics[width=17.0cm]{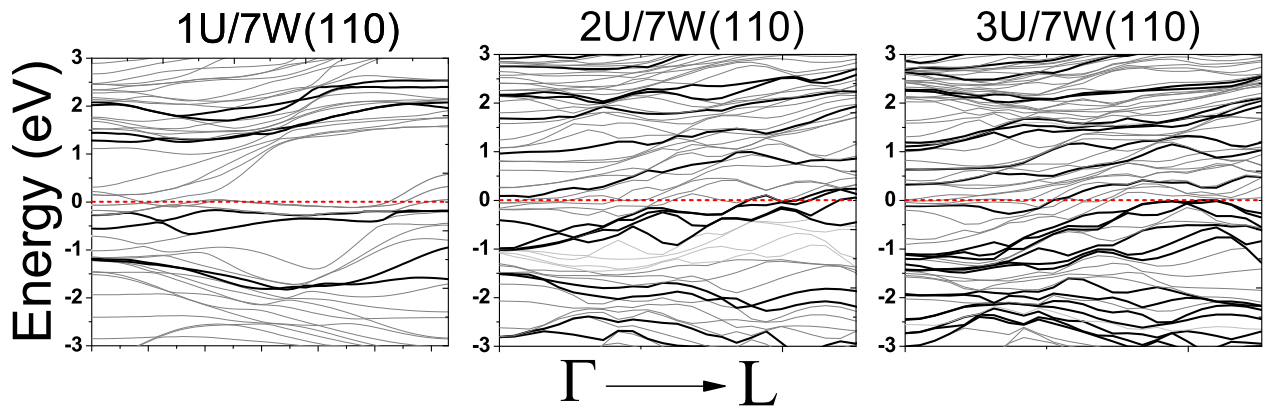}
\caption{(Color online) Band structures of 1U/7W(110), 2U/7W(110) and 3U/7W(110) systems.
The light lines indicate the total and the dark lines indicate
the uranium f-orbital bands.}
\label{fig4}
\end{figure*}

\begin{table}
 \begin{center}
  \caption{Total MM in $\mu_B$ per primitive unit cell
   within the PBE-GGA for three cases when the U1 is (i) in its relaxed equilibrium position, d$^{sur}_{U1-W1}$=2.526~\AA,
   (ii) moved upwards by 0.50~\AA, d$^{sur}_{U1-W1}$=3.026~\AA, (iii) moved upwards by 1.00~\AA,
   d$^{sur}_{U1-W1}$=3.526~\AA, with respect to the W1, while the other ILDs given in Tab.~\ref{tab1}
   were kept fixed.}\label{tab5}
  \begin{ruledtabular}
     \begin{tabular}{lcccccc}
                     &d$^{sur}_{U1-W1}$+0~\AA&&d$^{sur}_{U1-W1}$+0.5~\AA&&d$^{sur}_{U1-W1}$+1~\AA&\\
      \hline
      1U/7W(110)     &4.56&& 5.39 &&5.54\\
      2U/7W(110)     &0.45&& 0.53 &&0.57\\
      3U/7W(110)     &0.28&& 0.33 &&0.35\\
               \end{tabular}
  \end{ruledtabular}
 \end{center}
\end{table}

In Sec.~\ref{Comparison} we related the larger MM of the freestanding $\alpha$-U(001) than that of the $\alpha$-U/W(110) to the contracted vertical IL U-U distances of the $\alpha$-U/W(110) compared to that of the freestanding $\alpha$-U(001). However, there we did not discuss the role of the tungsten substrate. The effect of tungsten substrate may become more important when we compare the MM of the freestanding 1 ML $\alpha$-U(001) with that of 1U/7W(110). The MM per U atom within the GGA+SO for the freestanding 1 ML $\alpha$-U(001), 2.074 $\mu_B$, is larger than that for the 1U/7W(110), 2.988/2 = 1.494 $\mu_B$, as shown in Tabs.~\ref{tab2} and ~\ref{tab3}. This is in the case that there is no other uranium layer over the first uranium layer for both 1 ML $\alpha$-U(001) and 1U/7W(110). Therefore, for these one ML U cases there are not any vertical U-U interactions. Thus, the vertical U-U interactions could not be the case for the contraction and as a result for the reduction of the MM of the 1U/7W(110) compared to that of the 1 ML $\alpha$-U(001). In this section, we show that, in addition to the vertical IL U-U spacings, the in-plane U-U spacings may also affect the MM. In order to achieve this goal, we first discuss the proximity effect to elucidate the role of tungsten on its first uranium adatoms through the vertical W-U interaction at the interface of the 1U/7W(110) system. We then try to show the effect of tungsten through its imposed lattice mismatch on the in-plane U-U spacings. We discuss that not only the vertical IL U-U interactions, but also the in-plane U-U interactions can influence the MM.
To quantitatively elucidate the effect of the distance between the substrate and the film, we calculated, as presented in Tab.~\ref{tab5}, the total MM per primitive unit cell within the GGA when the U1 is vertically displaced 0.5~\AA~and 1.0~\AA~upwards with respect to its relaxed equilibrium position at d$^{sur}_{U1-W1}$ = 2.526~\AA. In the displacement procedure, it is assumed that apart from the d$^{sur}_{U1-W1}$ distance all of the other distances are kept fixed with no changes in their relaxed values as given in Tab.~\ref{tab1}. The result, as presented in Tab.~\ref{tab5}, shows that the total MMs increase as the U1 together with its fixed overlayers, if any, is moved upwards from its equilibrium position with respect to the fixed nonmagnetic 7W(110) substrate. The total MMs increase by $\sim$15\% when the layer is pulled up by 0.5~\AA. The percentage does not change significantly (more or less $\sim$21\%) when the layer is pulled up by 1.0~\AA~compared to the equilibrium position. Total MM was found\cite{Sto03} to be 3.27 $\mu_B$ per U for 1 ML $\alpha$-U within GGA. Our GGA result for the 1U/7W(110) is 4.56 $\mu_B$ per primitive unit cell and thereby 4.56/2 = 2.28 $\mu_B$ per U atom when U1 is placed at its equilibrium position. It increases from 2.28 $\mu_B$ to 5.54/2 = 2.77 $\mu_B$ as the overlayer is pulled away by 1~\AA~from the tungsten substrate. Thus the proximity effect of the substrate can cause our GGA result of 2.77 $\mu_B$ for the 1U/7W(110) to be closer to the GGA result\cite{Sto03} of 3.27 $\mu_B$ for the freestanding system compared to the value of 2.28 $\mu_B$ for 1U/7W(110) before its uranium overlayer is moved up. However, our result is still 15\% smaller than the freestanding result. The source of the later difference may not be only due to the proximity effect of the substrate, as further increase in vertical spacing of d$^{sur}_{U1-W1}$ does not affect the result substantially. Indeed, it may originate from the in-plane lattice mismatch effect of the substrate. The in-plane lattice mismatches can cause the lattice parameters of $\alpha$-U, as given in the text and Figs.~\ref{fig1}, to expand by 36.7\% along $\mathbf{a}$ axis and to contract by 45.1\% along $\mathbf{b}$ axis. Therefore, the difference may not be considered so large by accounting the large in-plane lattice mismatches. More precisely, the in-plane U-U distance between nearest neighbors for the case of the freestanding $\alpha$-U is $\sqrt{(2.842)^2+(5.780)^2}/2 \simeq 3.22$~\AA, which is larger than that for the $\alpha$-U/7W(110) system, i.e., $\sqrt{(4.492)^2+(3.176)^2}/2 \simeq 2.75$~\AA. For deriving the former value, we have used the lattice parameters of the $\alpha$-U, as shown in Fig.~\ref{fig1} (b), while for deriving the later value we have used the lattice parameters of the W(110) substrate, as shown in Fig.~\ref{fig1} (a). Although both of these  distances are smaller than the Hill limit, the in-plane U-U distance for the freestanding is larger and as a result closer to the Hill limit than that for the $\alpha$-U/W(110) system. Consequently, it seems to be reasonable to find larger MM for the freestanding case compared to the MM of the supported case by the tungsten substrate. In addition to the in-plane lattice mismatches and the proximity effects of the substrate, vertical lattice mismatch of the substrate can be also of significant importance in this 5f system. In contrast to the ultrathin 1U/7W(110) film, the 2U/7W(110) and 3U/W(110) TFs, as shown in Tab.~\ref{tab4}, do not manifest considerable magnetism even by moving upwards the U1 overlayer from the nonmagnetic substrate. This once more shows that the proximity effect of the substrate cannot be the only case here, since it has been already shown\cite{Sto03} that the freestanding 3 ML $\alpha$-U is a magnetic system. As discussed in Sec.~\ref{Rel}, the distance between uranium overlayers in 2U/7W(110) system is remarkably contracted by 19.1\%, as shown in Tab.~\ref{tab1}, compared to the bulk of $\alpha$-U. The contraction for the 3U/7W(110) is 15.7\% between U3 and U2 and 13.2\% between U2 and U1. The contractions can cause a stronger direct interaction between 5f states of two vertically adjacent uranium atoms. The stronger 5f-5f interaction causes the valence DOS to be more broadened. Total- or 5f-DOS of U1, as shown in Figs.~\ref{fig2}~(d) and (g), for U1/7W(110) shows a nearly narrow sharp peak in the vicinity of $E_F$. However, by adding more uranium overlayers the sharp narrow peak is more broadened, as can be seen from Figs.~\ref{fig2}~(e), (f), (h), (i) compared to Fig.~\ref{fig2}~(d) and (g). Hence, one may anticipate to observe strong reduction in magnetism from 1U/7W(110) to 2U/7W(110) or 3U/7W(110) in complete accordance with our result presented in Tabs.~\ref{tab3} and~\ref{tab5}. The comparison between Fig.~\ref{fig2}~(e) or (f) with Fig.~\ref{fig2}~(d) or between Fig.~\ref{fig2}~(h) or Fig.~\ref{fig2}~(i) with Fig.~\ref{fig2}~(g) corroborate the anticipation, since due to the direct interactions the DOSs are more symmetric and broadened. If we compare total- and f-DOSs of U1 with each other for the ultrathin U1/W(110) film, as shown in Fig.~\ref{fig2}~(d), we see that, in complete accord with experiments\cite{Rei85, Mol98, Ber04}, most of the total-DOS is dominated by the 5f states in the vicinity of the $E_F$. In order to authenticate the above discussion we have calculated the band structures of 1U/7W(110), 2U/7W(110) and 3U/7W(110) systems. Our results are shown in Fig.~\ref{fig4}. The calculated band structures show that the 5f U band character dominates in the vicinity of $E_F$ in agreement with experiment\cite{Rei85, Mol98, Ber04} and our calculated DOS. The comparison of band structures between 1U/W(110) and 2U/W(110) TFs shows that the degeneracy of 5f band is lifted by the vertical U-U interactions at -1 eV.  The removed degeneracy causes the band structures of the 2U/7W(110) to be more dominated in the vicinity of the $E_F$ by the 5f character of the uranium compared to that of the ultraTF of 1U/W(110). The number of lifted 5f bands increases in 3U/W(110), as shown in Fig.~\ref{fig4}, as the third uranium layer is deposited on the 2U/W(110) TFs. Such an increase in the number of splitted 5f bands in the 3U/W(110) can be taken as more interactions among 5f electrons of the uranium atoms in this system. This shows that strong interactions among 5f U electrons in uranium TFs in the presence of tungsten substrate may not allow the 3U/W(110) system to manifest magnetization at its surfaces. In conclusion the tungsten substrate can cause to destroy the appearance of magnetization on the hypothetical freestanding $\alpha$-U(001) TFs.

\section{Summary and Outlook}\label{Sum}
In this study, we considered the $\alpha$-structure for the uranium overlayers based on the recent experimental observations. We transformed the surface of the orthorhombic $\alpha$-uranium along the [001] direction, and found it to be similar to that of the bcc-tungsten substrate along the [110] direction. The surface structures of the W(110) and the transformed $\alpha$-U(001) are both fcc-like, as discussed in the text. We notice that the $\alpha$-U might be an ideal adsorbate for the W(110) tungsten taking vertical and in-plane lattice mismatches into consideration. This is in agreement with the recent experiment. We then simulated the $\alpha$-U/W(110) TFs and calculated its surface and electronic properties. The calculations were performed within the DFT employing the accurate FP-APW+lo method as embodied in the reliable WIEN2k code. Our result shows the proximity effect of the nonmagnetic tungsten substrate on the magnetism of uranium overlayers. In addition to the proximity effect, the imposed vertical and in-plane lattice mismatches of the W(110) substrate were found to be of significant importance. Our result predicts surface magnetism in the ultrathin 1U/7W(110) film. This result is in agreement with that of the freestanding 1ML $\alpha$-U(001). By adding more uranium layers on the ultrathin film, we found that the imposed surface magnetism disappears. This is in the case that for the freestanding $\alpha$-U(001) TFs, the magnetism is preserved but its magnitude decreases by adding one more uranium layer. The disappearance of the magnetism in 2U/7W(110) is attributed to the contraction of the uranium ILS, which can be originated from the W(110) substrate. The contraction can give rise to stronger direct 5f-5f interactions. Our calculation shows that the 5f U DOS is more broadened in
2U/7W(110) or 3U/7W(110) compared to the ultrathin 1U/7W(110) film. The broadening of the 5f DOS corroborates that the 5f states is more band-like for 3U/7W(110). This is in agreement with the experimentally founded dispersive behavior for the 5f states in U/W(110) TFs. Our result in agreement with experiment shows that the 5f character predominates the total DOS in U/W(110) TFs. According to our results, we anticipate further theoretical and experimental elaborations on the U/W(110) TFs taking the $\alpha$-structure for the adsorbate uranium into
consideration. For example, various orientations of the $\alpha$-U overlayers may improve the results, provided that less lattice mismatches can be found with the tungsten substrate. Theoretical investigation of hcp-U/W(110) TFs may also provide a significant contribution to the filed.

\acknowledgments This work, as a part of M.Z. thesis, is supported by the Office of Graduate Studies, University of Isfahan (UI),
Isfahan, Iran.

\end{document}